# Large (bi)polarons for novel energy-conversion and superconductivity

David Emin
Department of Physics and Astronomy
University of New Mexico
Albuquerque, New Mexico 87131, USA
emin@unm.edu

The last fifty years produced fundamental advances in understanding the formation, motion and interactions among (bi)polarons. Interference between the atomic displacement patterns of oppositely charged (bi)polarons generates a short-range repulsion which impedes their reconalmbination, significantly enhancing efficiencies of energy-conversion devices, e.g. perovskite solar cells. A large-(bi)polaron's atomic vibrations are softened by the relaxation of its self-trapped electronic carriers. Scattering of impinging ambient phonons by large-bipolarons' softened vibrations generates distinctive transport. Coherence between the softened atomic vibrations of large-bipolarons generates a phonon-assisted mutual attraction which facilitates their condensation into a liquid. A large-bipolaron liquid's superconducting Bose condensate is characterized by coherent zero-point atomic vibrations which decrease positron annihilation and increase ion channeling. Disorder can trigger electronic charge carriers' collapse into small polarons. The back-and-forth Arrhenius hopping of a small-polaron softens the associated atoms' vibrations thereby increasing their entropy. This effect explains the empirically observed Meyer-Neldel compensation effect. In particular, the temperature-independent factor of the jump-rate increases exponentially in proportion to its activation energy.

## 1. Introduction

I first met Ted Geballe when I spoke about small polarons at a Gordon Conference he and Al Clogston organized almost 50 years ago. Since that time much has been learned about polarons' formation, properties and interactions with one another. Here I will succinctly describe essential features of this polaron physics. I will stress the distinctive and useful properties of large-(bi)polarons that form in materials with especially displaceable ions as indicated by exceptionally large ratios of their static to high-frequency dielectric constants.

## 2. Large-polaron formation

An electronic charge carrier that interacts sufficiently strongly with the relatively heavy and slow-moving atoms of condensed matter becomes *self-trapped*, bound in the potential well produced by carrier-induced shifts of atoms' equilibrium positions.[1] In addition, relaxation of the self-trapped electronic carrier in response to displacements of atoms from these shifted equilibrium positions reduces the associated vibration frequencies.[2] A strong-coupling *polaron* refers to the composite quasi-particle comprising a self-trapped electronic carrier and the altered state of surrounding atoms.

The term *polaron* was introduced in anticipation of self-trapping primarily occurring through a charge carrier's Coulomb interactions with the displaceable ions of a *polar* (ionic) material. The displaceability of a material's ions is measured by the ratio of its static to high-frequency dielectric constants, $\varepsilon_0/\varepsilon_\infty$. In common covalent materials $\varepsilon_0/\varepsilon_\infty$ is only slight greater than 1. However, in classic ionic materials whose charge carriers form polarons, such as alkali halides, $\varepsilon_0/\varepsilon_\infty$ is almost 2. In the perovskite-based materials employed in novel solar cells and in novel superconductors, $\varepsilon_0/\varepsilon_\infty$ is at least 10.[3-8] Polaron formation and their phenomena are expected to be dramatic in materials with such huge values of $\varepsilon_0/\varepsilon_\infty$.

The adiabatic approach is generally utilized to address electronic and atomic motions in semiconductors. This method presumes that electrons readily adjust to the motions of relatively heavy and slow-moving atoms. The Emin-Holstein scaling approach addresses polaron formation in a continuum comprising an electron amongst displaceable atoms in the adiabatic *limit* in which vibrations are suppressed.[9] The system's eigenstates correspond to minima of the energy functional for the interacting system with respect to its electronic state's radius. The energy functional of a self-trapped electronic charge carrier having bandwidth parameter $T_e$ that interacts with (1) the atoms it envelopes via short-range interactions characterized by $V_s$ and (2) displaceable ions via long-range Coulomb interactions is:

$$E(R) = \frac{T_e}{R^2} - \frac{V_s}{2R^d} - \left(\frac{1}{\varepsilon_\infty} - \frac{1}{\varepsilon_0}\right)\frac{U}{2R}. \quad (1)$$

Here $U$ measures the Coulomb interaction between unscreened charges, $d$ denotes the electron's dimensionality and the electronic radius $R$ is dimensionless. The factor of two in the denominator of the potential-energy contributions presumes that electron-phonon interactions and strain energies respectively depend linearly and quadratically on atoms' displacements from their carrier-free equilibrium positions.

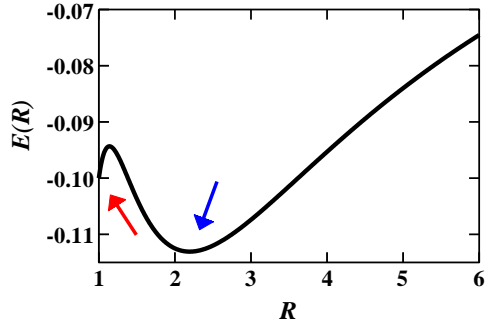

Fig. 1 The energy functional $E(R)$ for a self-trapped charge carrier with both short- and long-range components of its electron-lattice interaction is plotted versus its radius $R$. The metastable small-polaron and stable large-polaron minima are indicated by the red and blue arrows, respectively.

Most generally, as illustrated in Fig. 1, there are two distinct minima of $E(R)$ versus $R$ for a three-dimensional ($d = 3$) electronic carrier. The minimum at $R = 1$ corresponds to the self-trapped state collapsing to a single structural unit (atom, bond, ion or molecule) thereby forming a *small* polaron. The second minimum corresponds to the self-trapped carrier extending over multiple structural units thereby forming a *large* polaron. In the limit of an ideal covalent system, $\varepsilon_0 = \varepsilon_\infty$, the binding energy of this large-polaron solution vanishes as its radius increases to infinity. This unbound solution corresponds to the charge carrier remaining free.

### 3. Small-polaron formation

Disorder can readily induce a free carrier to collapse into a small polaron. In particular, amorphous covalent semiconductors are characterized by variations of their bond angles. The resulting large variations of the electronic transfer energies governing an electron and hole respectively passing among anti-bonding and bonding orbitals produces sites through which a free-carrier's passage is slow enough to induce its collapse into a small polaron. Then the impulse the carrier imparts to surrounding atoms exceeds their vibrational momentum.[10] To remain free, a carrier must avoid sites at which it would collapse into a small polaron. All told, the structural disorder of an amorphous covalent semiconductor reduces the electron-phonon coupling strength above which small-polaron formation occurs.[11]

A bound large polaron can be induced to collapse into a small polaron.[12] For example, a large-radius donor electron in the ferromagnetic semiconductor EuO abruptly collapses into small-radius states as the temperature is raised toward the ferromagnet's Curie temperature.[13] As illustrated in Fig. 2, this thermally induced collapse is driven by lowering the free energy associated with aligning the spin of the donor electron with those on Eu sites it encompasses. This collapse of donor states dramatically reduces their overlap from what it is when donor states have large radii. With sufficient donor densities (∼ 1%), their collapse suppresses impurity conduction among large-radius donors, reducing the electrical conductivity by about 15 orders of magnitude. Applying a magnetic field offsets the thermally induced misalignments of the ferromagnet's spins thereby raising the temperature of the donor-state collapse and that of the abrupt fall of the electrical conductivity.[14] In other words, there is a temperature domain within which imposing the magnetic field increases the conductivity. Such a huge negative magneto-resistance is very much larger than what is now termed CMR, colossal magnetoresistance.

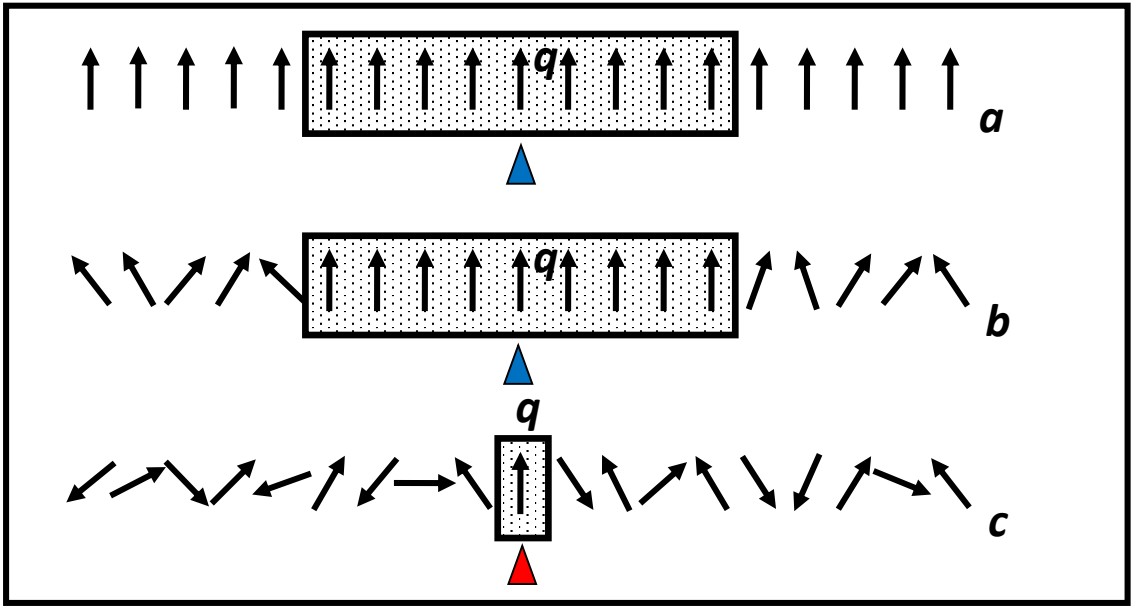

Fig. 2 The magnetic polaron comprising a self-trapped carrier of charge q and its ferromagnetically-aligned spins within a ferromagnet is depicted as contained within a rectangular box at three successively higher temperatures, $a \to b \to c$. The entropic contribution to the free energy from disordering spins drives the collapse of the polaron as the temperature is raised toward the Curie temperature.

### 4. Bipolaron formation

Electronic carriers can self-trap as singlet pairs to form bipolarons.[15-17] The energy functional for a singlet whose carriers share a common state $E_2(R)$ differs from that for a polaron $E(R)$ given in Eq. (1) in three respects: (1) the carriers' net confinement energy is doubled, (2) the potential energy contributions are quadrupled since each of the two carriers occupies a self-trapping potential that is doubled, and (3) the two electronic carriers repel one another through their mutual Coulomb repulsion. This bipolaron energy functional can be compared with that for two separated polarons:[2]

$$E_2(R) = \frac{2T_e}{R^2} - \frac{4V_s}{2R^d} - \left(\frac{1}{\varepsilon_\infty} - \frac{1}{\varepsilon_0}\right)\frac{4U}{2R} + \frac{U}{\varepsilon_\infty R} = 2E(R) + \frac{U}{\varepsilon_0 R} - \frac{V_s}{R^d}. \quad (2)$$

It is seen that (1) the long-range component of the electron-phonon interaction reduces the effective Coulomb repulsion between the pair by replacing $\varepsilon_\infty$ with $\varepsilon_0$ and (2) the short-range component of the electron-phonon interaction is needed to stabilize this bipolaron with respect to two separated polarons.

Anticipating this calculation being applied to the formation of large-bipolarons on $CuO_2$ layers in cuprate superconductors, consider $d = 2$. The energy of this two-dimensional large-bipolaron is just the minimum of $E_2(R)$ with respect to $R$. By comparison, the energy of two independent polarons, is just the minimum of energy functional $2E(R)$. Comparing these minima yields the conditions for a large bipolaron (1) being stable with respect to two separated large polarons and (2) not collapsing into a small bipolaron:[2,16,17]

$$1 > \frac{V_s}{T} > \frac{4(\varepsilon_0/\varepsilon_\infty) - 6}{(\varepsilon_0/\varepsilon_\infty)^2 - 2}. \quad (3)$$

The window within which these conditions can be fulfilled opens as $\varepsilon_0/\varepsilon_\infty$ increases above 2. Thus, large-bipolaron formation requires materials with especially displaceable ions, such as perovskite-based solids. This conclusion is unaltered upon generalizing this analysis to include correlation between the two states that comprise the large-bipolaron's singlet.

### 5. Properties of individual large-(bi)polarons

Motion of a self-trapped carrier requires significant motion of the associated atoms. As a result, polaron velocities are slower than the speed of sound. Furthermore, the effective mass of a coherently moving large-(bi)polaron is orders of magnitude greater than that of a conventional electronic carrier. In particular, the effective mass of a large (bi)polaron is $m_p \sim E_p/(\omega R_p)^2$, where $E_p$ and $R_p$ represent the large-(bi)polaron's binding energy and radius.[2,16]

A conventional fast and light electronic carrier is strongly scattered by relatively slow atomic vibrations. By contrast, the region of softened vibrations that surrounds a large (bi)polaron "reflects" the relatively fast ambient phonons that impinge upon it.[18]

Large (bi)polarons are most effectively scattered by very long-wavelength phonons, $> 2R_p$. In particular, the large-(bi)polaron scattering rate is the product of (1) the density of ambient phonons with wavelengths comparable to the large-(bi)polaron's spatial extent, (2) the interaction cross section and (3) the scattering induced change of (bi)polaron velocity:

$$\frac{1}{\tau_p} \sim \left[\frac{1}{R_p^d}\left(\frac{kT}{\hbar\omega}\right)\right][R_p^{d-1}]\left[\frac{2\hbar(\pi/R_p)}{m_p}\right] = \frac{2\pi kT}{m_p \omega R_p^2} \sim \left(\frac{2\pi kT}{E_p}\right)\omega. \quad (4)$$

Thus, the massive large (bi)polaron is very weakly scattered by ambient phonons to yield a scattering rate that is less than even the atomic vibration frequency $\omega$ since $kT \ll E_p$. This very weak scattering compensates the huge large-bipolaron mass to generate a mobility $\mu_p$ that is comparable to 1 cm$^2$/V-s at room temperature:

$$\mu_p = \frac{q\tau_p}{m_p} \sim \frac{q\omega R_p^2}{kT}, \quad (5)$$

where $q$ denotes the charge on the large (bi)polaron.

The strong phonon scattering of conventional electronic carriers and the weak phonon scattering of coherently moving large (bi)polarons both yield mobilities that decrease with rising temperature. To distinguish between these two situations, note that a carrier's mean-free-path $\Lambda$ must exceed its size (deBroglie wavelength) $\lambda$. Imposing this condition establishes that a carrier's minimum mobility is inversely proportional to its effective mass $m$: $\mu_{min} = (q/kT)<v\Lambda> \sim (q/kT)<(h/m\lambda)\Lambda> = qh/mkT$.[2,19] In particular, the minimum room-temperature mobility of a conventional carrier with an effective mass even as large as the free-electron mass is 300 cm$^2$/V-s. By contrast, since large (bi)polarons are massive, they can have room-temperature mobilities that are orders-of-magnitude smaller. For example, the room-temperature Hall mobilities of large polarons in alkali halides are well below 100 cm$^2$/V-s. Thus, carriers with room-temperature mobilities that are significantly smaller than 300 cm$^2$/V-s and fall with rising temperature are presumably large (bi)polarons.

Distinctively, the very weak scattering of coherently moving large (bi)polarons described in Eq. (4) limits the Drude contribution to their absorption spectra to *below* the phonon energy $\hbar\omega$.[20] In addition, a broad absorption band arising from exciting large (bi)polarons' self-trapped carriers from the potential well within which they are bound occurs at energies *above* $3E_p \gg \hbar\omega$.[20] Exciting self-trapped carriers to higher-lying bound states contributes some lower-energy absorptions. All told, large (bi)polarons are characterized by a *pseudo-gap* between their low-energy Drude absorptions and the high-energy absorptions of their self-trapped electronic carriers. This pseudo-gap opens with decreasing temperature as the Drude contribution shifts to lower energies.

## 6. Properties of individual small-(bi)polarons

The energy variation of a small-polaron as it moves between sites generally exceeds the corresponding transfer energy. As a result, small-polaron motion becomes incoherent. The corresponding motion it then best described as phonon-assisted hopping.[2]

Above a temperature that is a fraction of the corresponding phonon temperature, the small-polaron jump rate is Arrhenius with an activation energy greater than the characteristic phonon energy.[2] As the temperature is reduced and multi-phonon jump processes are progressively frozen out, the jump rate becomes non-Arrhenius with ever weakening temperature dependences. This non-Arrhenius behavior, which also occurs in crystals, is often mistakenly ascribed to the variable-range hopping that Mott suggested occurs between shallow impurities at very low temperatures.[21] At the lowest temperatures the jump rate for a hop upward in energy again becomes Arrhenius with an activation energy that just equals the difference of the polaron energy between the jump's final and initial sites. In this extreme low-temperature regime a jump downward in energy becomes temperature independent.[2,21]

Small-polaron hopping, unlike hopping between widely separated shallow impurities, is generally adiabatic: the jump's inter-site electronic transfer energy is large enough to enable the electronic carrier to adjust to energetic changes at the two sites.[2,22] As such, the pre-exponential factor in the high-temperature Arrhenius region becomes independent of the jump's inter-site transfer energy: $P \cong 1$ in $R_{hop}(s) = P(\omega/2\pi) \exp[-E_a(s)/kT]$. The range of jumps is then primarily limited by the increase of a hop's activation energy $E_a(s)$ with jump distance $s$.[2]

The activation energy for a high-temperature small-polaron hop is the minimum strain-energy required to bring the energy of the self-trapped electronic carrier at a jump's initial and final sites into coincidence. Often this energy cannot be dissipated as vibrational energy before a subsequent hop occurs. Successive hops are then no longer independent of one another.[23-25] Rather, small-polaron hops occur in flurries. In equilibrium, periods of rapid motion compensate for quiescent periods.[2] The transient relatively high mobility of injected carriers can persist for a considerable time before equilibration is established.[26]

The back-and-forth motion associated with a hop being followed by an immediate return hop softens the associated vibrations, thereby contributing to a jump's vibrational entropy.[27,28] Including this effect, the adiabatic jump rate becomes $R_{hop} = P(\omega/2\pi) \exp(E_a/t) \exp(-E_a/kT)$, where $t$ denotes the electronic transfer energy associated with moving between the two sites. This formula describes the empirically observed Meyer-Neldel compensation rule. The increase of the jump rate's temperature-independent factor upon increasing $E_a$ compensates for the decrease of the jump-rate's Arrhenius contribution.

Distinctively, the Hall Effect signs for small-polaron hopping are often anomalous.[29] Then a magnetic field deflects *n*-type small polarons in the same sense of a positively charged free carrier and *p*-type small polarons in the same sense as a negatively charged free carrier. Such Hall Effect sign anomalies are observed in most chalcogenide glasses as well as in doped amorphous Si, Ge and As.

Singlet small-bipolaron formation on a boron-carbide icosahedron is driven by its pair's small on-site Coulomb repulsion.[30,31] Low-temperature ESR and magnetic susceptibility measurements indicate singlet formation.[32] Nonetheless, the lowest activation energy occurs for a small-bipolaron's two carriers hopping as individuals.[33] Charge transport then predominantly occurs by breaking singlet pairs. Indeed, ESR measurements find a thermally activated spin density whose activation energy is close to that of the mobility.[34]

Polaron theory in general, and the theory of small-polaron hopping in particular, relies on the adiabatic approach.[2] This method presumes the carrier's mass to be very much smaller than those of the vibrating particles among which it diffuses. Nonetheless, light-interstitial diffusion in metals and other instances of atomic diffusion have been modelled in strict analogy with small-polaron hopping. This procedure encounters an obvious problem. In particular, the diffusing particle's sensitivity to host-atom's positions, its electron-phonon interaction, decreases with increases of its isotopic mass. As a result, strict small-polaron theory gives the activation energy for light-interstitial diffusion decreasing for increasing isotope mass. By contrast, the activation energy for hydrogen diffusion in metals is observed to increase with increasing isotope mass. A generalization of small-polaron theory having multiple bound states with differing inter-site tunneling rates resolves this discrepancy.[2,35]

## 7. Interactions between polarons

Interference between the atomic displacements induced by oppositely charged polarons produces an interaction between them.[2,22,36] Figure 3 schematically depicts the net energy of two oppositely charged polarons in a strongly ionic medium $E(s)$ as a function of their separation $s$. At large separations the net polaron energy approaches that of two independent oppositely charged polarons, $-(E_{p+} + E_{p-})$. As $s$ falls toward the sum of the two polaron radii, $R_{p+}$ and $R_{p-}$, constructive interference at intervening ions enhances the polarons' net binding energy. However, as $s$ is decreased further the net binding energy of the oppositely charged polarons falls rapidly as the Coulomb fields they generate at surrounding sites progressively cancel one another. This cancellation reduces carrier-induced displacements of surrounding ions. At $s = 0$ the oppositely charged carriers merge into an exciton of energy $E_{ex}$. With no net charge, the exciton lacks the Coulomb field needed to displace distant ions. Most significantly, when $\varepsilon_0 \gg \varepsilon_\infty$ interference between ionic displacements induces a repulsive interaction between oppositely charged polarons which dominates their mutual Coulomb attraction to yield a net short-range inter-polaron *repulsion*.

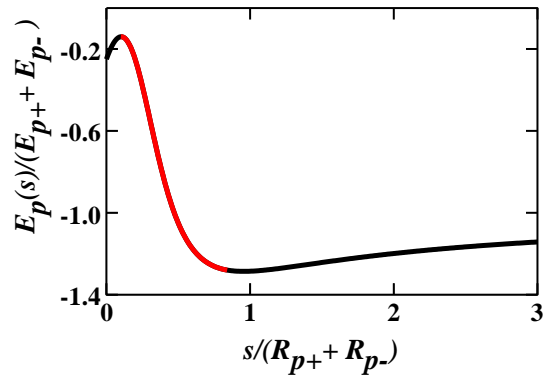

Fig. 3. The energy of two oppositely charged polarons $E_p(s)$ divided by $(E_{p+} + E_{p-})$, the sum of their individual binding energies, is plotted against their separation $s$ divided by $R_{p+}$ and $R_{p-}$, the sum of the two polaron radii. As the inter-polaron separation is decreased from infinity, displacements of the equilibrium positions of intervening ions induced by the two polarons increases causing their net energy to decrease. However, the net interference of the ionic displacements changes from being constructive to being destructive as the two polarons begin to overlap. The surrounding ions then increasingly see the two polarons as having no net charge. The two polarons ultimately collapse into an exciton when $s = 0$. The red portion of $E_p(s)$ versus $s$ indicates the region of the two polarons' mutual short-range repulsion.

The net repulsion between oppositely charged polarons in a strongly ionic medium, $\varepsilon_0 \gg \varepsilon_\infty$, suppresses their recombination.[36] This phenomenon enables using strongly ionic materials as the basis for novel solar cells in which optically generated carriers form polarons. Indeed, the recently discovered high-efficiency perovskite-based solar cells seem to exemplify this situation.[37] These metallic- and organometallic-halide perovskite materials have 1) extremely large values of $\varepsilon_0/\varepsilon_\infty$,[6-9] 2) moderate rrier mobilities that are comparable to those of alkali halides' large polarons,[38-42] 3) large-polaron-like absorption bands,[43] and 4) suppressed recombination.[44] Remarkably, the efficiencies of these novel perovskite solar cells, > 20%, are much larger than those of conventional silicon-based solar cells, ~10%. All told, conventional covalent-semiconductor solar cells separate their oppositely charged high-mobility photo-induced carriers quickly enough to limit their recombination. By contrast, novel solar cells employ strongly ionic solids whose moderate-mobility oppositely charged large polarons experience short-range mutual repulsions which suppress their recombination.

Solar-cell-type devices also can be used to convert the power of nuclear decays into electrical power.[45] In these devices bombardment with products of nuclear decays produce electron-hole pairs whose separation at *p-n* junctions generate electricity. Such devices powered by beta-particle bombardment are termed beta cells. However, radiation damage from bombarding conventional Si and Ge solar cells with beta particles from $^{90}$Sr severely reduces performance within only a day.[46] As a result, beta cells using conventional semiconductors only employ very weak sources to generate miniscule electrical power, ~1μW.

Icosahedral-boron semiconductors are unusual materials based on covalent networks of linked twelve-atom boron icosahedra.[30,31] Three distinctive features of some icosahedral boron semiconductors may facilitate producing long-lasting high-efficiency beta cells that generate significant electrical powers.[47] First, atomic displacements induced by their bombardments with very energetic electrons spontaneously *self-heals*.[48] An electron-deficient boron icosahedron then retains an electron from an exiting boron atom thereby rendering it a cation. Self-healing occurs as small boron cations are drawn to structurally stable "degraded" icosahedra by the extra electron each garners. Second, low-mobility *p*-type electronic transport generally occurs on the electron-deficient boron network.[49] Nonetheless, materials based on the $MgAlB_{14}$ structure manifest moderate-mobility *n*-type electronic transport once sufficient numbers of electrons are donated to the boron network from metal atoms that partially occupy large extra-network open regions.[50] Third, the displaceable cations between which these electrons move facilitate their forming large polarons.[9] The short-range repulsion which suppresses polaron recombination can then enhance a beta cell's energy conversion efficiency.[36]

The intensity of the near-steady beta emission from $^{90}$Sr incorporated within $SrTiO_3$, 1000 W/m$^2$, is much greater than the average of the extremely variable terrestrial solar intensity (100 W/m$^2$ in Britain and 200 W/m$^2$ in the desert southwest, e.g. Albuquerque).[45,51] Moreover, $^{90}$Sr, is a cheap and abundant (~5%) constituent of reactor waste whose bremsstrahlung can be readily shielded.[45] Thus, $^{90}$Sr beta cells based on $MgAlB_{14}$ promise significant reliable stand-alone electric power.[47] A beta cell's efficiency should be compared with that of a $^{90}$Sr-based 500 W RTG, Radioisotopic Thermoelectric Generator, ~7%.[45]

## 8. Collective motion and large-bipolaron superconductivity

A large bipolaron, like a large polaron, moves coherently with a mobility that increases with decreasing temperature.[2] By contrast, small polarons and bipolarons generally move incoherently via thermally assisted hopping. Thus, *large* bipolarons are plausible constituents of a collective superconducting state.

A bipolaron has a singlet pair of electronic carriers bound in the ground-state of its self-trapping potential well. A grander polaron would have additional carriers bound within its self-trapping potential well. However, such entities are destabilized by the Pauli principle's requiring that additional carriers occupy excited states of the self-trapping potential well. In particular, the entity containing four self-trapped electronic carriers is unstable with respect to separating into two large bipolarons. Thus, there is a short-range repulsion between large bipolarons that is akin to that between $^4$He atoms.[2]

An intermediate-range attraction between large bipolarons is generated by their atoms' vibrations.[52,53] Relaxation of large bipolarons' self-trapped electronic carriers reduces the stiffness constants for displacing atoms from their carrier-induced equilibrium positions. As a result, increasing the large-bipolaron density lowers the solid's net phonon energy. The associated phonon-mediated attraction between two large bipolarons is of the order of a phonon energy.

This phonon-mediated attraction will dominate the long-range mutual Coulomb repulsion between large bipolarons when, as in cuprate superconductors, the material's static dielectric constant $\varepsilon_0$ is large enough. Then combining this intermediate-range phonon-mediated attraction and the short-range repulsion enables large-bipolarons to condense into a liquid as the temperature is reduced. Thus, large bipolarons' condensation into a liquid is analogous to that by which $^4$He atoms condense into liquid $^4$He.[52,53]

A large-bipolaron liquid is regarded as a liquid of charged bosons. Its Bose condensation occurs when the thermal energy falls below a value comparable to the charged liquid's plasma energy, $\propto (m\varepsilon_0)^{-1/2}$.[2,17,54] This plasma energy is exceptionally small, less than a phonon energy, because large bipolarons are very massive and their mutual Coulomb interactions are reduced by the solids' huge static dielectric constants.[3-5]

The Bose condensation produces a finite occupation of the liquid's ground-state. The finite value of the energy of a charged-boson-liquid's excitations $E(p)$ at zero momentum, $p = 0$, its plasma energy, insures its satisfying Landau's condition for its ground-state exhibiting resistance-less flow: $E(p)/p > 0$ as $p \to 0$.[2] In addition to the onset of superconductivity, the Bose condensation signals the onset of its superfluid's homogeneous charge distribution with synchronously vibrating zero-point atomic vibrations. As such, positron annihilation associated with the self-trapped electronic charge and the scattering of channeled ions are progressively reduced with the growth of the homogeneous condensate.[55-57]

Distinctive features characterize a large-bipolaron superconductor. 1) The electronic chemical potential is pinned by bipolarons' self-trapped carriers since they occupy the bound states their very presence induces. In addition, the spread of carriers' electronic energies narrows as they progressively join the homogeneous superconducting ground-state. 2) The excitations of a large-bipolaron liquid are primarily scattered by very long wavelength acoustic phonons to generate a low-temperature mobility that is proportional to $1/T$. As such, the normal-state resistivity for a temperature-independent density of carriers introduced by doping is proportional to $T$.[58] 3) A pseudo-gap exists in the absorption spectrum of a large-bipolaron liquid.[20,58] With decreasing temperature, the very weak scattering of the excitations of a large-bipolaron liquid constrain its Drude absorption to progressively lower frequencies below the phonon frequency, $< \hbar\omega$. Ionizations of large bipolarons' self-trapped carriers primarily occur above the phonon frequency, $> \hbar\omega$. 4) The condensate of a large-bipolaron liquid cannot exhibit a proximity effect with a metal since it cannot exist there. However, this condensate can pass through a material whose dielectric constants permit large-bipolarons and their condensates.

I envision a $p$-type large bipolaron being formed on a $CuO_2$ plane of a cuprate when two electrons are removed from four $O^{2-}$ anions bounded by four $Cu^{2+}$ cations.[59] The collateral outward relaxation of the four Cu cations is accompanied by their each taking an electron from an inward relaxing oxygen anion: 2 holes + $4O^{2-}$ + $4Cu^{+2}$ → $(O_4)^{2-}$ + $4Cu^{+1}$. Thus, the formation of a $p$-type bipolaron on a $CuO_2$ layer removes four copper spins.

Three unusual observations of cuprate superconductors are readily explained in terms of large-bipolaron superconductivity.

    1) The loss of an unpaired spin on each of the four outwardly relaxing copper cations removes its spin entropy. Thus, the Seebeck coefficient, the entropy transported by a charge carrier divided by its charge, for hole-like large-bipolarons garners the contribution $-4(k/2e) \ln 2$ as the temperature is raised above the superconducting transition temperature toward that characterizing the paramagnetic regime.[60-64]

    2) The self-trapped electronic singlet at the core of a large bipolaron is distributed over the out-of-plane oxygen orbitals of the four inwardly relaxing oxygen atoms. There are four radial normal-mode optic

vibrations of these four oxygen atoms. The highest-frequency mode is the breathing mode having *s*-symmetry. The two degenerate intermediate-frequency modes have *p*-symmetry and the lowest-frequency mode has *d*-symmetry. Since the amplitudes of the zero-point vibrations of these modes are inversely proportional to the square roots of their frequencies, the *d*-symmetry vibrations have the largest amplitudes. Thus, the synchronous zero-point vibrations of the collective ground-state has primarily *d*-symmetry.[59] The self-trapped electrons occupying these four vibrating oxygen atoms also reflect this symmetry.[59]

3) The ground-state of a collection of bipolarons will no longer remain fluid if it solidifies by ordering commensurate with the underlying lattice. Thus, superconductivity will be lost in tetragonal hole-doped $La_2CuO_4$ when the hole doping of the $CuO_2$ plane becomes 2/(5x5), 2/(4x4) and 2/(3x3). The two extreme compositions are close to the limiting compositions of the superconducting domain whereas the middle composition occurs in the middle of the superconducting region at 1/8, where superconducting is observed to vanish.[65]